\documentclass{an}

\usepackage{graphicx,bm,color,url}
\graphicspath{{./fig/}{./figs/}{./png/}}

\newcommand{\EQ}{\begin{equation}}
\newcommand{\EN}{\end{equation}}
\newcommand{\EQA}{\begin{eqnarray}}
\newcommand{\ENA}{\end{eqnarray}}

\newcommand{\Eq}[1]{Equation~(\ref{#1})}

\newcommand{\Sec}[1]{Section~\ref{#1}}

\newcommand{\Fig}[1]{Figure~\ref{#1}}

\newcommand{\Figs}[2]{Figures~\ref{#1} and \ref{#2}}

\newcommand{\Tab}[1]{Table~\ref{#1}}

\newcommand{\meanrho}{\overline{\rho}}

{}
{}
{}

{}
{}
\newcommand{\meanEMF}{\overline{\mbox{\boldmath ${\cal E}$}}{}}{}
{}
{}
{}
{}
{}
\newcommand{\meanAA}{\overline{\mbox{\boldmath $A$}}{}}{}
\newcommand{\meanBB}{\overline{\mbox{\boldmath $B$}}{}}{}
{}
{}
{}
{}
{}
{}
{}
{}
\newcommand{\meanJJ}{\overline{\mbox{\boldmath $J$}}{}}{}
{}
\newcommand{\meanUU}{\overline{\bm{U}}}

{}
{}
{}

\newcommand{\meanU}{\overline{U}}

{}

{}
\newcommand{\emf}{{\cal E}}{}

\newcommand{\alphaK}{\alpha_{\rm K}}
\newcommand{\alphaM}{\alpha_{\rm M}}

\newcommand{\nullvector}{{\bf0}}

\def\bb{\bm{b}}
\newcommand{\BB}{\mbox{\boldmath $B$} {}}

\newcommand{\jj}{\mbox{\boldmath $j$} {}}

\newcommand{\nab}{\mbox{\boldmath $\nabla$} {}}

\newcommand{\ii}{{\rm i}}

\newcommand{\pa}{{\partial}}

\newcommand{\const}{{\rm const}  {}}

\def\St{\mbox{\rm St}}

\def\Rm{\mbox{\rm Re}_M}

\def\Hp{H_{\rm p}}

\def\kf{k_{\rm f}}

\def\Brms{B_{\rm rms}}

\def\urms{u_{\rm rms}}

\def\etat{\eta_{\rm t}}
\def\etatz{\eta_{\rm t0}}

\def\Beq{B_{\rm eq}}

\def\onethird{{\textstyle{1\over3}}}

\newcommand{\km}{\,{\rm km}}

\newcommand{\Mm}{\,{\rm Mm}}

\newcommand{\yapj}[3]{ #1, {\it ApJ,} {#2}, #3}
\newcommand{\yapjl}[3]{ #1, {\it ApJL,} {#2}, #3}
\newcommand{\yapjs}[3]{ #1, {\it ApJS,} {#2}, #3}
\newcommand{\yan}[3]{ #1, {\it Astron.\ Nachr.,} {#2}, #3}
\newcommand{\yana}[3]{ #1, {\it A\&A,} {#2}, #3}
\newcommand{\yproc}[5]{ #1, in {#3}, ed.\ #4 (#5), #2}
\newcommand{\ygafd}[3]{ #1, {\it Geophys.\ Astrophys.\ Fluid Dyn.,} {#2}, #3}

\newcommand{\ypf}[3]{ #1, {\it Phys.\ Fluids,} {#2}, #3}
\newcommand{\ysov}[3]{ #1, {\it Sov.\ Astron.,} {#2}, #3}
\newcommand{\yprl}[3]{ #1, {\it Phys.\ Rev.\ Lett.,} {#2}, #3}
\newcommand{\ymn}[3]{ #1, {\it MNRAS,} {#2}, #3}
\newcommand{\ysci}[3]{ #1, {\it Science,} {#2}, #3}
\newcommand{\ypre}[3]{ #1, {\it Phys.\ Rev.\ E,} {#2}, #3}
\newcommand{\yjour}[4]{ #1, {\it #2}, {#3}, #4}
\newcommand{\ybook}[3]{ #1, {\it #2} (#3)}

\begin{document}
\title{Strong nonlocality variations in a spherical mean-field dynamo}
\author{A. Brandenburg\inst{1,2,3,4} \& P. Chatterjee\inst{5}}
\institute{
Laboratory for Atmospheric and Space Physics, University of Colorado, Boulder, CO 80303, USA
\and
JILA \& Dept.\ of Astrophysical and Planetary Sciences, University of Colorado, Boulder, CO 80303, USA
\and
Nordita, KTH Royal Institute of Technology and Stockholm University, 10691 Stockholm, Sweden
\and
Department of Astronomy, Stockholm University, 10691 Stockholm, Sweden
\and
Indian Institute of Astrophysics, II Block Koramangala, Bengaluru-560034, India
}
\date{\today,~ $ $Revision: 1.56 $ $}

\abstract{
To explain the large-scale magnetic field of the Sun and other bodies,
mean-field dynamo theory is commonly applied where one solves the averaged
equations for the mean magnetic field.
However, the standard approach breaks down when the scale of the
turbulent eddies becomes comparable to the scale of the variations of the
mean magnetic field.
Models showing sharp magnetic field structures have therefore been
regarded as unreliable.
Our aim is to look for new effects that occur when we relax the restrictions
of the standard approach, which becomes particularly important at the
bottom of the convection zone where the size of the turbulent eddies is
comparable to the depth of the convection zone itself.
We approximate the underlying integro-differential equation by a partial
differential equation corresponding to a reaction-diffusion type
equation for the mean electromotive force, making an approach that is
nonlocal in space and time feasible under conditions where spherical
geometry and nonlinearity are included.
In agreement with earlier findings, spatio-temporal nonlocality lowers the
excitation conditions of the dynamo.
Sharp structures are now found to be absent.
However, in the surface layers the field remains similar to before.
}

\maketitle

\section{Introduction}

The solar convection zone (CZ) is strongly stratified not just in the sense
that the density contrast is enormous, but also in the sense that the
pressure scale height $\Hp$ varies strongly: from $150\km$ at the top
of the CZ to about $50\Mm$ at the bottom (Stix 2002).
Given that the size of the turbulent eddies is expected to be
proportional to $\Hp$, this change in $\Hp$ corresponds to a 300 fold
increase in the typical scale of the turbulent eddies.
It is unclear whether this change in the turbulence properties can
just be translated to a corresponding position-dependent change of the
related turbulent transport coefficients (e.g., turbulent diffusivity
and $\alpha$ effect), as is usually assumed (e.g., Pipin 2008, 2017;
Pipin \& Kosovichev 2011), or whether one should expect some more subtle
effects as a result of strong stratification.
In the former case, turbulent transport can be described by just a
local Fickian diffusion law where the turbulent flux
of a passive scalar, for example, is proportional to the gradient of the mean
concentration and a diffusion coefficient that is given by the product
of the local values of turbulent velocity and correlation length
(Parker 1979; R\"udiger \& Hollerbach 2004).
The latter is usually assumed to be a multiple of the local pressure scale
height, and the turbulent velocity can be estimated from stellar
mixing length theory (Krivodubskii 1984).
The alternative to the local Fickian diffusion law is a nonlocal
formulation, in which the turbulent flux of passive scalar concentration
is given by a convolution with an integral kernel (R\"adler 1976).
Such a formulation might have unexplored effects when applied to
solar dynamo theory, where the turbulent concentration flux in the
Fickian diffusion law corresponds to the mean electromotive force
in the mean-field induction equation.
It is therefore necessary to investigate the effects of a strong scale
height change in a system resembling the solar dynamo,
but that is simple enough so that one has a chance
to isolate the consequences of individual effects.

In the standard mean-field prescription in terms of $\alpha$ effect
and turbulent diffusivity $\etat$, the mean electromotive force
$\meanEMF$ is given by (Krause \& R\"adler 1980)
\EQ
\meanEMF=\alpha\meanBB-\etat\mu_0\meanJJ,
\label{meanEMF}
\EN
where anisotropies have been ignored.
Here $\meanBB$ is the mean magnetic field, $\meanJJ=\nab\times\BB/\mu_0$
is the mean current density, and $\mu_0$ is the vacuum permeability.
However, \Eq{meanEMF}
becomes invalid when the mean field shows variations on short time and
length scales (Brandenburg et al.\ 2008a, Hubbard \& Brandenburg 2009).
In that case the multiplications with coefficients $\alpha$ and $\etat$
must be replaced by a convolutions with integral kernels  $\hat\alpha$
and $\hat\etat$.
In practice such an approach is cumbersome, because one has to store
the values of the mean fields at all past times at all positions.
An easier way to deal with this is to turn the underlying
integro-differential equation into a partial differential equation.
This can be done in sufficiently simple cases.
An example that has previously been explored in this context is
the telegraph equation approach, where one 
replaces $\meanEMF$ by $(1+\tau\partial_t)\meanEMF$
(Blackman \& Field 2002, 2003).
Here $t$ is time and $\tau$ is some relevant relaxation time
that is of the order of the turnover time of the turbulence; see
Brandenburg et al.\ (2004).
In spectral space, this corresponds to an integral kernel of the
form $(1-\ii\omega\tau)^{-1}$, where $\omega$ is the frequency;
see Hubbard \& Brandenburg (2009).
The spatial part of the integral kernel is known to be a Lorentzian
in spectral space (Brandenburg et al.\ 2008a), i.e.,
$(1+k^2/k_{\cal E}^2)^{-1}$,
where $k_{\cal E}$ is a relevant wavenumber that is expected
to be proportional to the wavenumber $\kf$ of the energy-carrying eddies
with a scale factor $a_{\cal E}=\kf/k_{\cal E}$.
This means that
$\meanEMF$ should be replaced by $(1-\ell^2\nabla^2)\meanEMF$.
Here $k$ is the wavenumber.
Recent work by Rheinhardt \& Brandenburg (2012) has shown that the
combination of both spatial and temporal effects at the same time yields
\EQ
\left(1+\tau{\partial\over\partial t}-\ell^2\nabla^2\right)
\meanEMF=\alpha\meanBB-\etat\mu_0\meanJJ,
\label{Nonlocal}
\EN
where $\ell\equiv k_{\cal E}^{-1}$ characterizes the length scale
on which nonlocality becomes important.
This equation has to be solved simultaneously with the usual
mean-field dynamo equation.
Obviously, this nonlocal formulation in space and time becomes equal
to the local one in the limit $\tau\to0$ and $\ell\to0$.
When applied to the Sun, we must expect the $\tau$ and $\ell$
terms to become important at the bottom of the CZ
where the turnover time and the correlation length are large.
When $\ell$ becomes large, the effect of the mean magnetic
field on $\meanEMF$ becomes reduced by what corresponds to a Lorentzian
$1/(1+\ell^2k^2)$ in Fourier space,
i.e., not only the $\alpha$ effect but also turbulent diffusivity
becomes progressively weaker in deeper layers of the Sun.

The aim of this paper is to assess the significance of
a strong vertical variation of the degree of nonlocality
in space and time on dynamo models that have stratification and
rotation profiles similar to those expected for the Sun.
We adopt a simple mixing length prescription through which rms velocity
as well as correlation length and correlation time depend in power law
fashion on the depth below the surface.

\section{Aspects of the model}

In this section, we motivate a number of aspects that should,
on physical grounds, be included in a solar mean-field model.
We begin by presenting first the basic equations, and then discuss
several profile functions that characterize a model of the solar dynamo.

\subsection{Dynamo equations}

In this paper, we ignore temporal changes of the mean flow
and consider only the changes of the mean magnetic field $\meanBB$,
whose evolution is given by
\EQ
{\partial\meanAA/\partial t}=\meanUU\times\meanBB+\meanEMF-\eta\mu_0\meanJJ,
\EN
where $\meanUU$ is a prescribed mean flow, $\meanBB=\nab\times\meanAA$
is the mean magnetic field expressed in terms of the vector potential
$\meanAA$, and $\mu_0$ is the vacuum permeability.
The mean flow includes the differential rotation, i.e.,
$\meanU_\phi=r\sin\theta\Omega(r,\theta)$, and in general also
meridional circulation, which is here, however, neglected.
Meridional circulation is particularly important in models with separated
induction zones (Choudhuri et al.\ 1995; Dikpati \& Charbonneau 1999;
Chatterjee et al.\ 2004).
However, in models with overlapping induction zones, as in the present
case, meridional circulation usually just increases the critical dynamo
number, but does not change significantly the shape of the eigenfunction
(R\"adler 1986).

\subsection{Nonlocality in earlier work}

Previous studies using the test-field method applied to the steady state
($\omega=0$) suggest that $\ell=a_\eta/\kf$,
where $a_\eta$ is a dimensionless
parameter that is usually in the range 0.5--1
(Brandenburg et al.\ 2008a, 2009; Mitra et al.\ 2009), but it can
be as low as 0.2 if there is shear (Madarassy \& Brandenburg 2010).
Furthermore, in the unsteady case, the test-field method suggests that
$\St=\tau\urms\kf$ is in the range 1.4--2.0 (Hubbard \& Brandenburg 2009).
Writing now \Eq{Nonlocal} as an evolution equation, we have
\EQ
{\partial\meanEMF\over\partial t}=
\tau^{-1}\left(\alpha\meanBB-\etat\mu_0\meanJJ-\meanEMF\right)
+\eta_{\cal E}\nabla^2\meanEMF,
\label{NonlocalEvol}
\EN
where $\eta_{\cal E}=a_{\cal E}^2\etat$ is a diffusion coefficient in the
evolution equation for $\meanEMF$, which is expected to scale like the
turbulent magnetic diffusivity $\etat\approx\onethird\tau\urms^2$.
Assuming furthermore that $\tau=\St/\urms\kf$, where $\St$ is the
Strouhal number (expected to be of the order of unity; see
Sur et al.\ 2008), we have $a_{\cal E}=3a_{\cal E}^2/\St$.

In this paper we examine the effects of nonlocality in a spherical
shell dynamo.
Recent work using simultaneous nonlocality in space and time has shown
that $\St=0.6$--1.2 and $a_{\cal E}=0.6$--0.8 for $\Rm\approx60$
(Rheinhardt \& Brandenburg 2012), corresponding to $a_{\cal E}=1.5$--3.

\subsection{Nonlinearity}
\label{Nonlinearity}

It should be noted that in the present approach the incorporation of
nonlinear feedbacks does not require any special consideration.
Both dynamical and algebraic quenching (see Brandenburg \& Subramanian
2005a for a review) can be applied in the usual sense.
This is remarkable, because the original formalism in terms of
integral kernels is a standard concept in linear response theory
and thus not readily applicable to the nonlinear regime; see the
discussion in Rheinhardt \& Brandenburg (2012).

At the simplest level we adopt just algebraic $\alpha$ quenching, by
which the local value of $\alpha$ is suppressed locally proportional to a
function
\begin{equation}
f_\alpha(\meanBB)={1\over1+Q_\alpha\meanBB^2/\Beq^2},
\end{equation}
where $\Beq=\sqrt{\mu_0\rho}\,\urms$ is the equipartition field strength.
In addition, we adopt the dynamical quenching model, in which we solve
an evolution equation for a magnetic contribution to the $\alpha$ effect,
i.e.\ $\alpha$ is the sum of kinetic and magnetic $\alpha$ effects, 
$\alpha=\alphaK+\alphaM$, where
$\alphaM=(\tau/3\meanrho)\,\overline{\jj\cdot\bb}$ is proportional to the
current helicity, which, in the isotropic case, is proportional to the
magnetic helicity of the small-scale field.
The magnetic $\alpha$ effect, in turn,
obeys an evolution equation that must be solved simultaneously.
The equation for the evolution of $\alpha_{\rm M}$ is given by
(e.g., Subramanian \& Brandenburg 2006)
\begin{equation}
\label{eq:alphaeq}
\frac{\pa \alpha_{\rm M}}{\pa t} = -2\eta_{\rm t} k_{\rm f}^2
\left(\frac{\overline{\vec{\mathcal{E}}}\cdot\overline{\vec{B}}}{B_{\rm eq}^2}
+\frac{\alpha_{\rm M}}{R_{\rm m}}\right)-\vec\nabla\cdot \overline{\vec{F}}_{\alpha},
\end{equation}
where $\overline{\vec{F}}_{\alpha}=-\kappa_\alpha\nab\alpha_{\rm M}$
is the turbulent--diffusive magnetic helicity flux, which was found
to be the most important contribution (Hubbard \& Brandenburg 2011, 2012;
Del Sordo et al.\ 2013).
Here $\kappa_\alpha$ is a turbulent diffusion coefficient for the flux.
In the following we assume $\kappa_\alpha=\etat$ as a good approximation.

Chatterjee et al.\ (2011) have solved \Eq{eq:alphaeq} along with the
mean field induction equation for an $\alpha\Omega$ interface dynamo in
a spherical shell.
They found that the time-latitude plot for the toroidal field shows signatures
of what they called `secondary dynamos' (see their Figs.~9 and 10). 
This phenomenon occurred both for vanishing magnetic helicity flux
as well as for fluxes that were `supercritical' in the sense that the
fluxes alone would drive a dynamo of the type proposed by Vishniac \&
Cho (2001); see Brandenburg \& Subramanian (2005b) for a mean-field
model of such a dynamo.
These magnetic helicity-driven dynamos were suspected to be
excited due to the lack of nonlocality in their model.
In this paper, we also run a similar nonlinear model
along with the equation for $\pa\meanEMF/\pa t$. 

In practice, we combine both algebraic and dynamic $\alpha$ quenching
and write
\begin{equation}
\alpha=f_\alpha(\meanBB)\,(\alphaK+\alphaM),
\end{equation}
Thus, the function $f_\alpha$ provides the ultimate limitation to $\alpha$
in the sense that $\alpha\to0$ for strong magnetic fields.

\subsection{Model stratification}

Instead of using a realistic stratification from a solar model,
we felt it more insightful to use a stratification based on simple
approximations such as a polytropic one and to get in this way
the radial dependence of density $\rho$.
Furthermore, according to mixing length theory (Vitense 1953; Brandenburg
2016), the convective flux is, to a good approximation, equal to
$\rho\urms^3$, where $\urms$ is the rms velocity of the turbulence.
This relation fails near the bottom of the solar CZ,
because there the radiative flux begins to dominate, so $\urms$
would be overestimated.
On the other hand, over sufficiently large ranges in radius $r$, it is not
the flux $F$ that is constant, but the luminosity $4\pi r^2 F$,
so $F$ should really be larger toward the bottom of the CZ,
so this would underestimate $\urms$.
Here we ignore the two opposing effects and assume that they
cancel each other to some extent.

We adopt an adiabatic stratification where the sum of local enthalpy,
$h$ ($=c_p T$ for a perfect gas with specific heat at constant
pressure $c_p$) and the gravitational potential $\Phi$
($=-GM/r$ for a point mass $M$, or $gz$ for plane-parallel layer
with constant gravitational acceleration $g$) are constant.
We assume that the mixing length $\ell$ is proportional to the
pressure scale height, $\Hp=|\nab\ln p|^{-1}$.
We assume a polytropic index of $n=3/2$, so $\rho\sim\ell^{3/2}$,
where $\ell$ is the mixing length, which is assumed to be proportional to
the depth below the surface, i.e., $\ell=R-r\equiv\kf^{-1}$
(cf.\ Canuto \& Mazzitelli 1991).

Using $\rho u^3=\const$, we have for the rms velocity the scaling
\EQ
\urms\sim\kf^{1/2},
\EN
which has its maximum at the surface.
This implies that the correlation or turnover time has the scaling
\EQ
\tau\sim(\urms\kf)^{-1}\sim\kf^{-3/2},
\EN
which has a maximum at the bottom of the CZ,
and similarly also the turbulent diffusivity $\etat$ and the $\alpha$ effect,
\EQ
\etat\sim\tau\urms^2\sim\kf^{-1/2},\quad
\alpha\sim\Omega\ell\sim\kf^{-1}.
\label{eq:alpha}
\EN
The local values of $\alpha$ and $\etat$ would therefore reach a maximum
at the bottom of the CZ (Krivodubskii 1984; Brandenburg \&
Tuominen 1988).
To avoid finite values of $\alpha$ and $\etat$ at $r=r_{\rm b}$,
we multiply $\kf^{-1}$ by an additional
profile function so that $\alpha$ goes smoothly to 0 at $r=r_{\rm b}$.
Furthermore, both the memory term, $\tau\omega_{\rm cyc}\sim\kf^{-3/2}$,
and the nonlocal term, $(\kf R)^{-2}\sim\kf^{-2}$, are maximum at the bottom.
This makes the dynamo locally less efficient in the lower layers.

The $\alpha$ effect changes sign about the equator and is proportional
to $\cos\theta$, while $\etat$ is independent of $\theta$, so we write
\EQ
\alpha(r,\theta)={\alpha_0 g(r)\cos\theta\over\kf(r)R},\quad
\etat(r)={\etatz g(r)\over(\kf(r)R)^{1/2}}.
\EN
Here $\alpha_0$ and $\etatz$ are coefficients that can be combined into
a dynamo number,
\EQ
C_\alpha=\alpha_0 R/\etatz.
\EN
In addition to a profile for $\etat$, we also adopt a profile for the
microscopic magnetic diffusivity $\eta$, which is assumed constant
everywhere except in the outer parts where $\eta/\etatz\to1$ to mimic
a potential (current-free) magnetic field.
Unlike the usual case of a local model, where only the combination
$\eta+\etat$ enters, the two are now different in that $\eta$ enters
directly in the induction equation for the magnetic field while $\etat$
enters in the evolution equation for the mean electromotive force.

\begin{figure}
\includegraphics[width=\columnwidth]{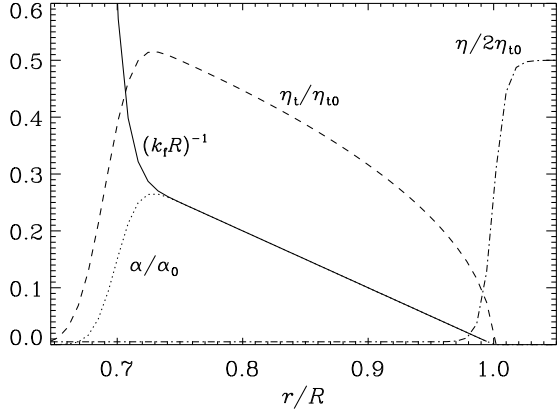}\caption{
Profiles of $(\kf R)^{-1}\equiv1-\ell/R$, $\alpha/\alpha_0$ and $\etat/\etatz$.
}\label{pkf_prof}
\end{figure}

In \Fig{pkf_prof} we plot radial profiles of $\kf^{-1}$, as well as
$\alpha/\alpha_0$, $\etat/\etatz$, and $\eta/\etatz$.

\begin{figure}[t!]\begin{center}
\includegraphics[width=\columnwidth]{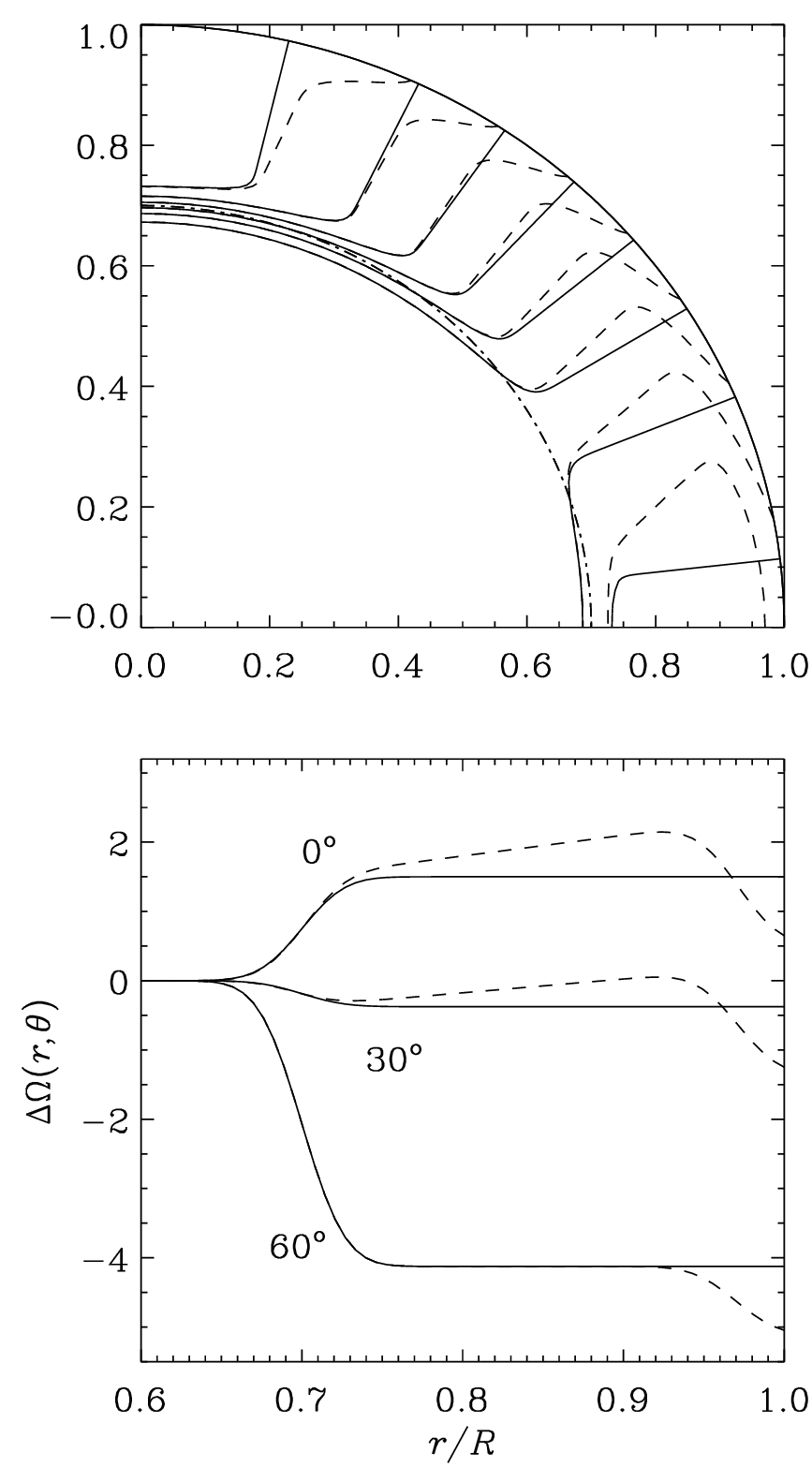}
\end{center}\caption[]{
Contours of differential rotation (upper panel)
and radial profiles of angular velocity at three
different latitudes (lower panel) from the simple
model of \Eq{DiffRot} with $c_{\Omega2}=0$
(spoke-like contours; solid lines) and $c_{\Omega2}=1$
(tilted contours with NSSL; dashed lines).
}\label{pdiffrot}\end{figure}

\subsection{Differential rotation profile}

We use a solar-like profile as a superposition of
contributions from the tachocline (TC), a small positive radial
differential rotation in the CZ, and a sharp negative
differential rotation in the near-surface shear layer (NSSL), i.e.,
\EQ
\Omega=\Omega_0\left[\Omega_{\rm TC}
+c_{\Omega2}\left(\Omega_{\rm CZ}+\Omega_{\rm NSSL}\right)\right],
\label{DiffRot}
\EN
where
\EQ
\Omega_{\rm TC}=-\Theta^+(r-r_{\rm b})
C_2^{3/2}(\cos\theta)
\EN
models the TC (the minus sign corresponds to equatorial acceleration),
\EQ
\Omega_{\rm NSSL}=-\Theta^+(r-r_{\rm t})
\EN
the NSSL, and
\EQ
\Omega_{\rm CZ}=(r-r_{\rm b})\,
\Theta^+(r-r_{\rm b})\,\Theta^-(r-r_{\rm t})
\left(3-4\cos^2\theta\right)
\EN
the interior region in the CZ.
Here,
\EQ
C_2^{3/2}(\cos\theta)
=-{P_3^1(\cos\theta)\over\sin\theta}
={3\over2}\left(5\cos^2\theta-1\right)
\EN
is the Gegenbauer polynomial of degree 3/2 and $P_3^1$ is the associated
Legendre polynomial of degree 3 and order 1.
In \Fig{pdiffrot} we compare contours of differential rotation
with just the TC contribution ($c_{\Omega2}=0$) and
with that using all contributions ($c_{\Omega2}=1$).
In the following we quantify the strength of differential rotation
by the nondimensional quantity $C_\Omega=\Omega_0 R^2/\etatz$.

We use the
{\sc Pencil Code}\footnote{\url{http://github.com/pencil-code}}
(revision 19,388 or later).
We use spherical polar coordinates, $(r,\theta,\phi)$.
We solve the equations in two dimensions ($\partial/\partial\phi=0$)
in $r_{\rm i}\leq r\leq r_{\rm o}$, which is large enough to encompass
the bottom of the CZ ($r=r_{\rm b}$) and the outer radius
of the sphere ($r=R$).
In the following, length is usually expressed in units of $R$.

\subsection{Model parameters}

The model geometry is defined by the following choice:
$r_{\rm i}=0.55$, $r_{\rm b}=0.7$, $R=1$, and $r_{\rm o}=1.05$.
We restrict ourselves to solving the equations in one quadrant
of the meridional plane ($0\leq\theta\leq\pi/2$, and adopt a
condition at the equator that selects only solutions that are antisymmetric
about the equatorial plane.

Unless specified differently, we adopt for all radial profiles
a width of $w=0.02$, which can still be reasonably well resolved with
just 64 mesh points in the radial direction.
In the latitudinal direction we take 96 mesh points. 

In all calculations, we make the $\alpha\Omega$ approximation,
i.e., we assume that the $\alpha$ tensor is finite only in its $\phi\phi$
component, so we can neglect the $\alpha$ effect in the generation of
the toroidal field in comparison with the differential rotation.
This means that the dynamo onset is only determined by one dynamo number,
namely $D=C_\alpha C_\Omega$.
This has the advantage that there is one parameter less to consider,
although it is somewhat unrealistic.

\begin{table}[t!]\caption{
Summary of runs mentioned in the text.
Nonlinear runs are marked by nlin and figure numbers
are given in the last column.
}\vspace{12pt}\centerline{\begin{tabular}{lccrccccc}
Run & $\St$ & $a_\emf$ & $D\;$ & $c_{\Omega2}$ & nlin &
$T_{\rm cyc}$ & $\Brms$ & Fig(s)\\
\hline
K1N & 1 & 1 &$\! 500\!$&  1 & no  & 0.526 & --- & \ref{d64_U3e4a} \\
K1L & 0 & 0 &$\!4000\!$&  1 & no  & 0.108 & --- & \ref{64_U3e4b}  \\
K0N & 1 & 1 &$\! 500\!$&  0 & no  & 0.531 & --- & \ref{d64_S3e4a} \\
K0L & 0 & 0 &$\!4000\!$&  0 & no  & 0.108 & --- & \ref{64_S3e4b}  \\
Q1N & 1 & 1 &$\! 100\!$&  1 & yes & 0.539 & 21.9& \ref{pb6_d64x96_U3e4aM6}, \ref{pbut_d64x96_U3e4aM6} \\
Q1L & 0 & 0 &$\!2500\!$&  1 & yes & 0.289 & 0.44& \ref{pb6_64x96_U3e4aM5}, \ref{pbut_64x96_U3e4aM5} \\
\label{Tsum}\end{tabular}}\end{table}

\section{Results}

In the following, we vary the parameters 
$\St$, $a_\emf$, $D$, and $c_{\Omega2}$.
Our runs are summarized in \Tab{Tsum}.

\subsection{Kinematic runs with NSSL}

\begin{figure}[t!]\begin{center}
\includegraphics[width=\columnwidth]{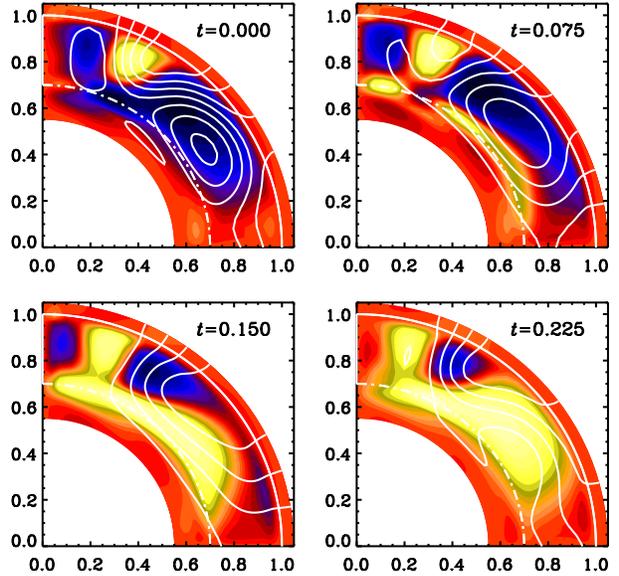}
\end{center}\caption[]{
Field lines in the meridional plane together with a color-coded
representation of the toroidal field (dark/blue shades indicate
negative values and light/yellow shades positive values).
Evolution of the field structure for Run~K1N with NSSL
and $D=500$ (slightly supercritical, oscillatory), using the
$\partial\meanEMF/\partial t$ equation with $\St=1$ and $a_{\cal E}=1$.
}\label{d64_U3e4a}\end{figure}

In \Figs{d64_U3e4a}{64_U3e4b} we compare nonlocal models (with
$\partial\meanEMF/\partial t$ equation included) and local ones (without
$\partial\meanEMF/\partial t$ equation) by showing field lines in the
meridional plane together with a color-coded representation of the
toroidal field.
Note that with the $\partial\meanEMF/\partial t$ equation, the
cycle frequency is about 5 times lower than without this equation
($\St=a_{\cal E}=0$),
and also the excitation condition is about 5 times lower.
This is in agreement with similar results in Cartesian geometry
(Rheinhardt \& Brandenburg 2012).

\begin{figure}[t!]\begin{center}
\includegraphics[width=\columnwidth]{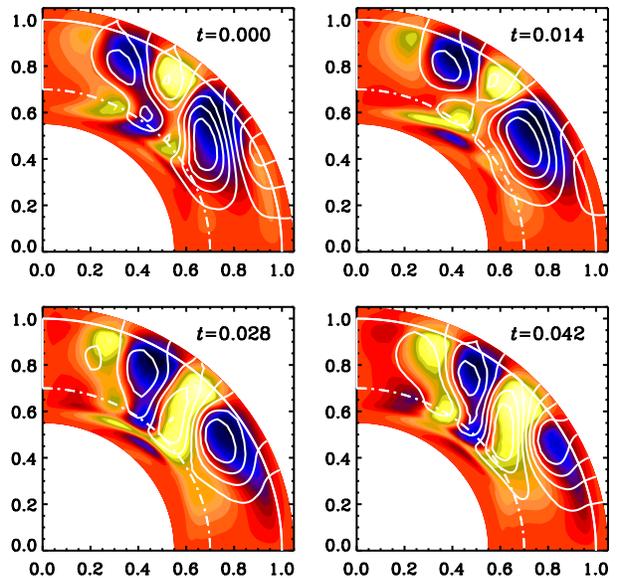}
\end{center}\caption[]{
Like \Fig{d64_U3e4a}, but for Run~K1L without the
$\partial\meanEMF/\partial t$ equation and $D=4000$
(slightly supercritical, oscillatory), using a local model,
i.e., $\St=a_{\cal E}=0$.
}\label{64_U3e4b}\end{figure}

\subsection{Kinematic runs without NSSL}

\begin{figure}[t!]\begin{center}
\includegraphics[width=\columnwidth]{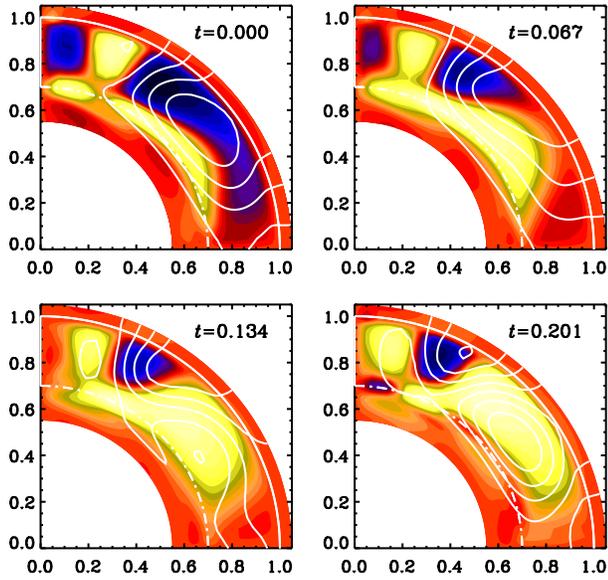}
\end{center}\caption[]{
Like \Fig{d64_U3e4a}, but for Run~K0N with $c_{\Omega2}=0$.
Note that the time $t=0$ corresponds here to a similar moment
at $t=0.075$ in \Fig{d64_U3e4a}.
Thus, the last time at $t=0.201$ in the present figure is
close to the negative field at $t=0$ in \Fig{d64_U3e4a}.
}\label{d64_S3e4a}\end{figure}

Let us now compare with the corresponding cases where the NSSL is omitted
(i.e., $c_{\Omega2}=0$).
In \Figs{d64_S3e4a}{64_S3e4b} we show again meridional cross-sections of the
magnetic fields for the same two cases as in \Figs{d64_U3e4a}{64_U3e4b}.
Note also that the magnetic field distribution is almost unchanged,
regardless of the absence or presence of the NSSL.
On the contrary, in the case
without the $\partial\meanEMF/\partial t$ equation,
the rms magnetic field strength actually
increases by about 25\% when $c_{\Omega2}=0$,
while there is no difference when the
$\partial\meanEMF/\partial t$ equation is included.
(The models presented in \Figs{d64_S3e4a}{64_S3e4b} are, however,
linear, so no rms value is given here.)

\begin{figure}[t!]\begin{center}
\includegraphics[width=\columnwidth]{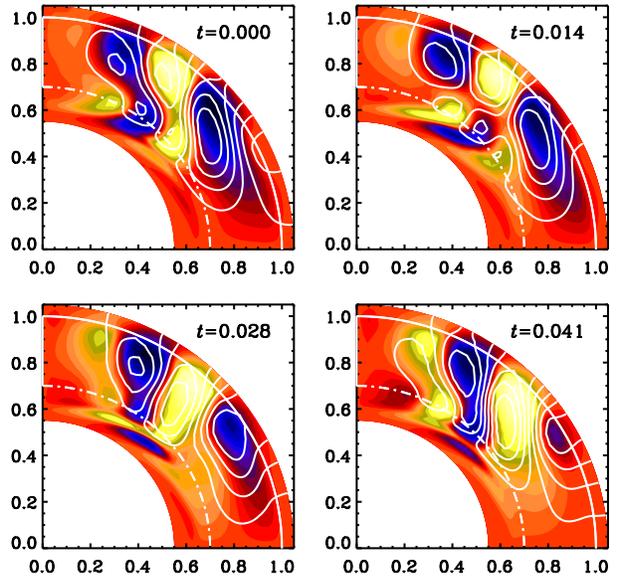}
\end{center}\caption[]{
Like \Fig{64_U3e4b}, but for Run~K0L with $c_{\Omega2}=0$.
Most of the small-scale structures appear near the bottom of the
CZ, so the effect of the  NSSL in \Fig{64_U3e4b}
is not so strong.
}\label{64_S3e4b}\end{figure}

\begin{figure}[t!]\begin{center}
\includegraphics[width=\columnwidth]{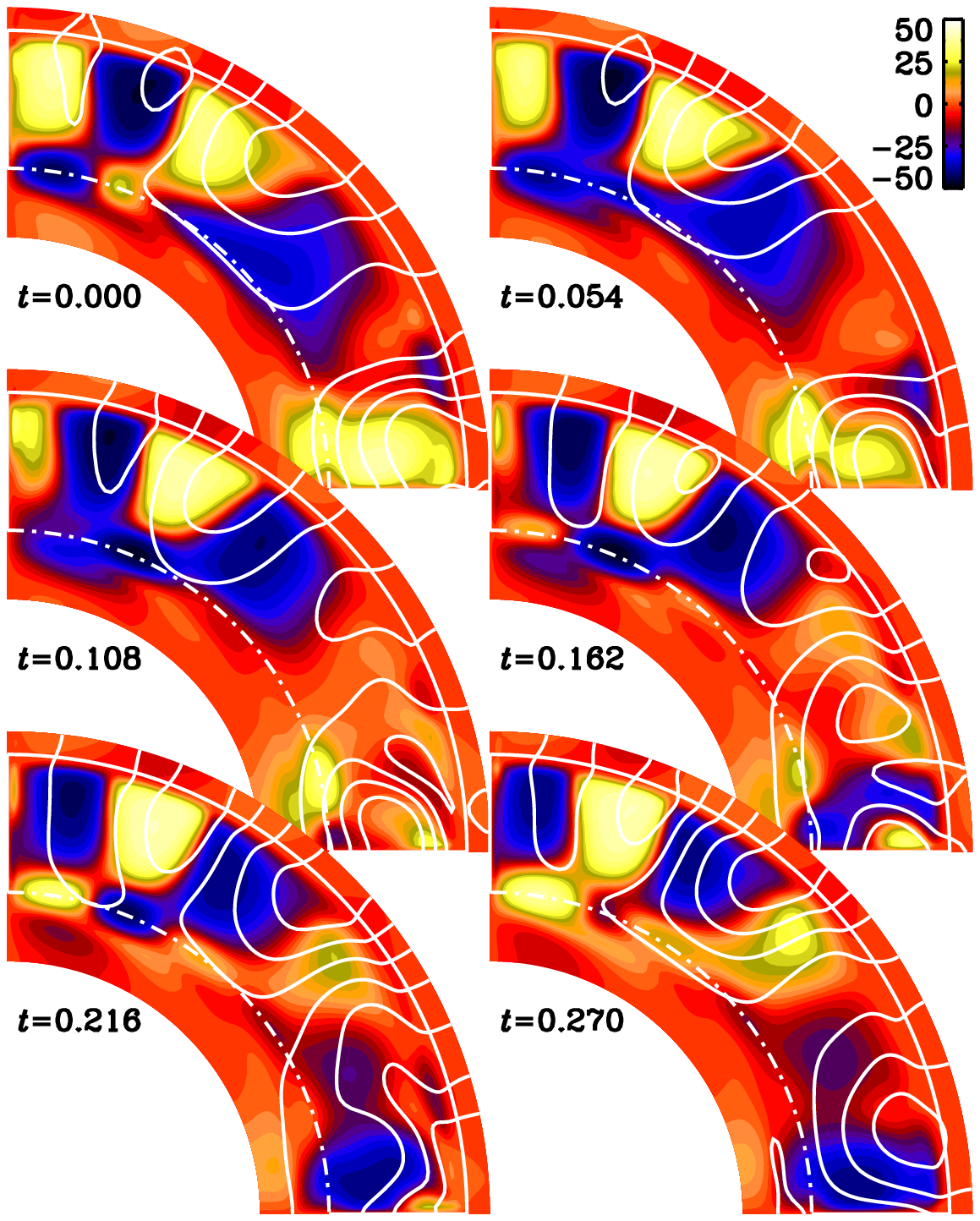}
\end{center}\caption[]{
Meridional cross-sections of the magnetic field of Run~Q1N
with $D=100$, $\St=1$, and $a_{\cal E}=1$.
}\label{pb6_d64x96_U3e4aM6}\end{figure}

\begin{figure}[t!]\begin{center}
\includegraphics[width=\columnwidth]{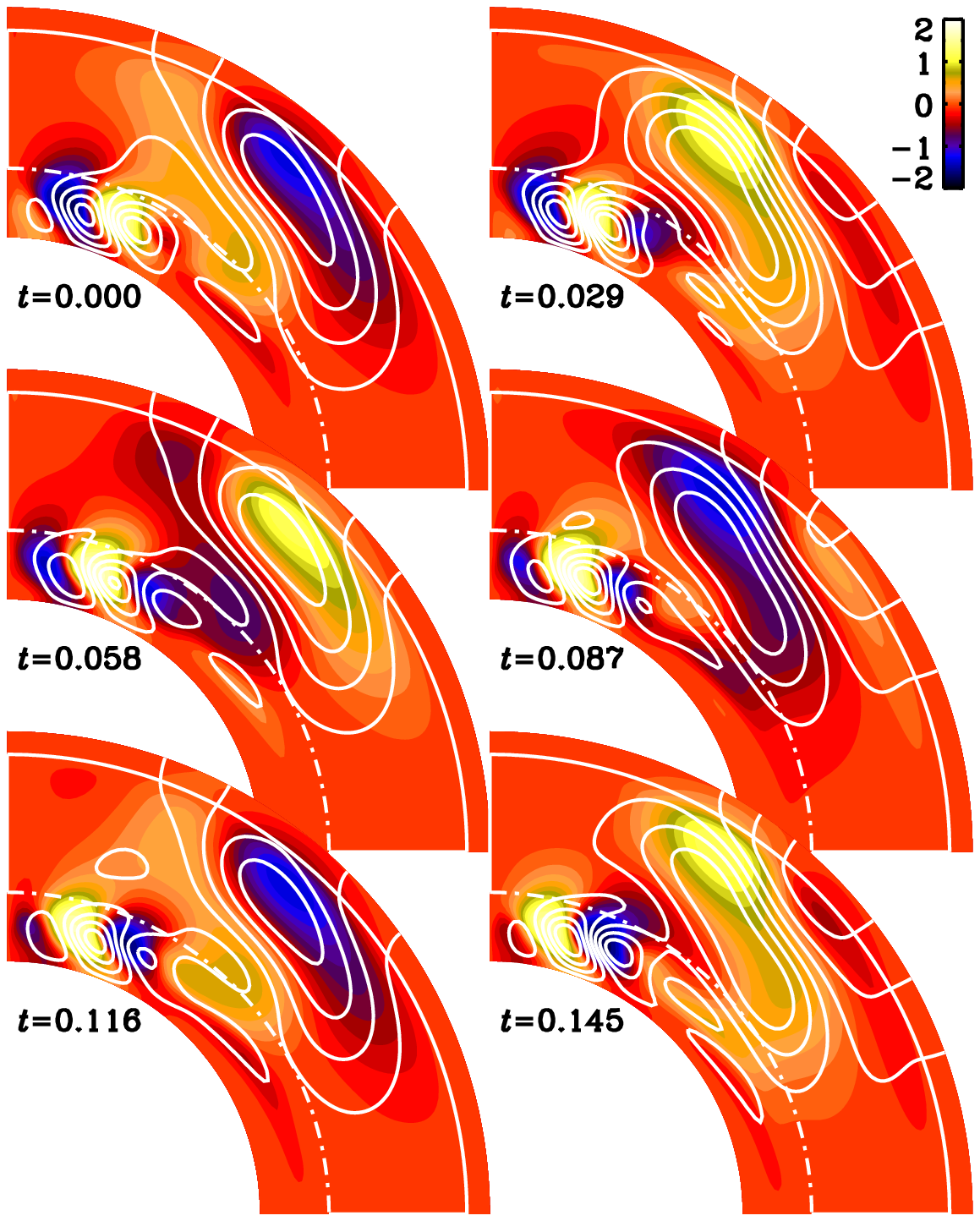}
\end{center}\caption[]{
Meridional cross-sections of the magnetic field of Run~Q1L
with $D=2500$, $\St=0$, and $a_{\cal E}=0$.
}\label{pb6_64x96_U3e4aM5}\end{figure}

\begin{figure}[t!]\begin{center}
\includegraphics[width=\columnwidth]{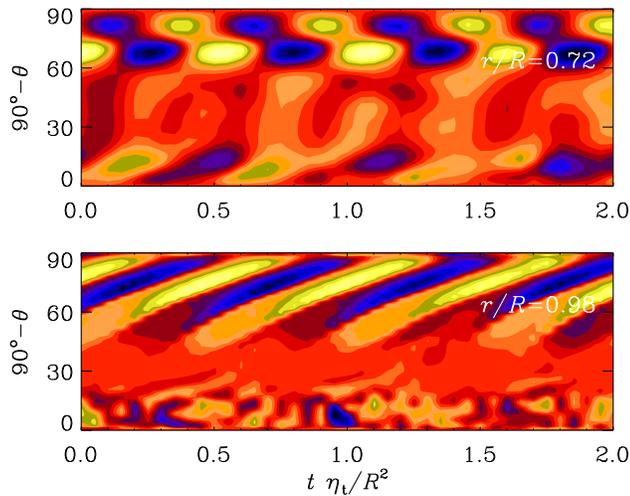}
\end{center}\caption[]{
Butterfly diagrams at $r/R=0.72$ (top) and $0.98$ (bottom)
for Run~Q1N, i.e., the same run as in \Fig{pb6_d64x96_U3e4aM6}
with dynamical quenching.
}\label{pbut_d64x96_U3e4aM6}\end{figure}

\begin{figure}[t!]\begin{center}
\includegraphics[width=\columnwidth]{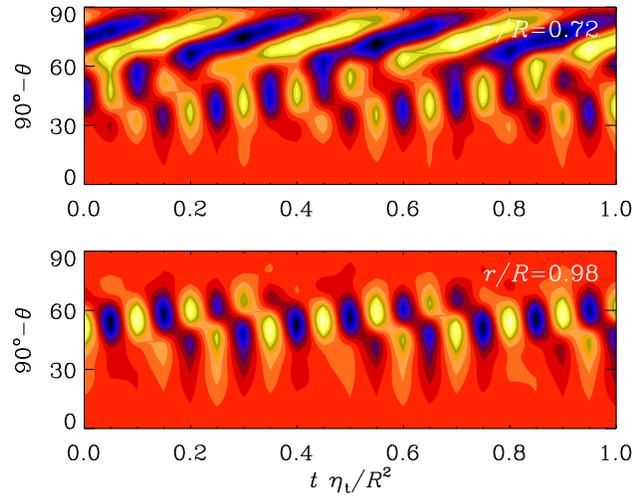}
\end{center}\caption[]{
Like \Fig{64_U3e4b}, but with dynamical quenching.
Butterfly diagram for Run~Q1L.
}\label{pbut_64x96_U3e4aM5}\end{figure}

Generally, the local models have smaller-scale structures at
the bottom of the layer, including a detailed radial dependence
in the lower overshoot layer.
Clearly, any mean-field structure on scales comparable to or below
the scale of the turbulent eddies must be suspect.
As expected, such structures are indeed absent in the more realistic
nonlocal treatment, where the magnetic field is now much smoother
in the deeper layers and there is no rapid field variation in the
overshoot layer.

An interesting aspect of the models is that without the
$\partial\meanEMF/\partial t$ equation, the time
variation is no longer sinusoidal, but there
appear different frequencies that cannot be
realistic, because the high frequencies disappear
when nonlocality is included.

\subsection{Nonlinear runs}
\label{NonlinearRuns}

The nonlocal approach presented here is readily applied to the
nonlinear case; see the discussion in \Sec{Nonlinearity}.
Such models are shown in \Figs{pb6_d64x96_U3e4aM6}{pb6_64x96_U3e4aM5}
for the cases with and without nonlocality.
Again, the main difference between the two models is in the presence of
small-scale structures at the bottom of the CZ.
As a result, the overall magnetic field appears much smoother
in \Fig{pb6_d64x96_U3e4aM6} than in \Fig{pb6_64x96_U3e4aM5}.
As in the kinematic case, the field is strongest at high latitudes, but
the nonlinear model now allows smaller-scale structures at low latitudes
(\Fig{pb6_d64x96_U3e4aM6}) that are not observed in the kinematic case
(\Fig{d64_U3e4a}).

Finally, we compare for the two cases in
\Figs{pbut_d64x96_U3e4aM6}{pbut_64x96_U3e4aM5}, respectively,
butterfly diagrams at the bottom of the CZ
(upper panels, $r/R=0.72$) and the top (bottom panels, $r/R=0.98$).
It is now clear that artificial small-scale structures are now produced
in the local model, especially at the surface (\Fig{pbut_64x96_U3e4aM5}).

\subsection{Catastrophic quenching}

\begin{figure}[t!]\begin{center}
\includegraphics[width=\columnwidth]{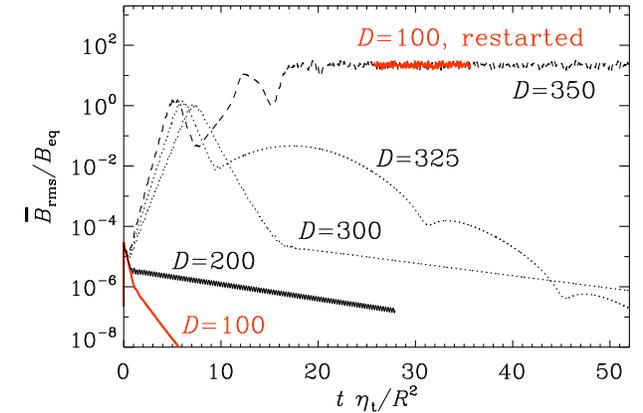}
\end{center}\caption[]{
Time series of runs for different values of $D$.
The run with $D=350$ is supercritical, while all other
runs ($D\leq325$) are subcritical.
Two realizations of Run~Q1N with $D=100$ are shown in red;
the upper one has been restarted from the run with $D=350$
at $t\etat/R^2\approx25$ and has a nearly unchanged rms
magnetic field.
}\label{pcomp}\end{figure}

The dynamical quenching runs discussed in \Sec{NonlinearRuns} have all
been restarted from an earlier run.
However, it turns out that their bifurcations from the trivial solution
($\BB=\nullvector$) are subcritical, i.e., they are finite amplitude
solutions and cannot be obtained from just an infinitesimal weak seed
magnetic field.
To demonstrate this, we show in \Fig{pcomp} the time series of runs for
different values of $D$.

For $D=325$ and below, all runs decay due to catastrophic quenching,
because their magnetic helicity flux is too weak, and so the resulting
$\alphaM$ profile never reaches the distribution found for
a saturated dynamo.
Only for a strong initial magnetic field is one able to arrive at a
regime of strong supercritical magnetic helicity fluxes.
This idea was first proposed by Vishniac \& Cho (2001) and later confirmed
with one-dimensional mean-field simulations by Brandenburg \& Subramanian
(2005b), who used a prescription for the magnetic helicity fluxes from
Vishniac \& Cho (2001).
However, as shown in Hubbard \& Brandenburg (2011), this contribution to
the magnetic helicity fluxes cannot be confirmed by direct numerical
simulations, where a turbulent--diffusive magnetic helicity fluxes was
found to operate instead.

\section{Conclusions}

This paper has demonstrated the great ease with which nonlocality in space
and time can be implemented in a mean-field model in spherical coordinates.
Unlike earlier work, we have here assumed that the nonlocality
parameters $\tau$ and $\ell$ are functions of position.
With this it is clear that we have entered a more speculative regime.
Indeed, simulations have so far only been carried out in the case where
the turbulence is statistically steady and homogeneous, so that $\tau$
and $\ell$ could be obtained as constant parameters over the whole domain
(Rheinhardt \& Brandenburg 2012).

The present results appear quite plausible in many respects.
In particular, while with conventional models small-scale
structures appear at the bottom of the CZ,
they now have been washed out when nonlocality is enabled.
This is reasonable because such structures were always regarded as artifacts.
One such example where this caveat was discussed was given by
Chatterjee et al.\ (2011); see their Figures~10 and 11. 
Moreover, in addition to allowing for nonlocality, our approach
also allows for nonlinearity in a straightforward fashion, as was
already pointed out by Rheinhardt \& Brandenburg (2012).

Interestingly, \cite{HRY16} showed that the magnetic energies in their
dynamo simulations first decreased and then increased with increasing
the Reynolds number from 30 to 3000, i.e., the dynamo was first
got quenched and then recovered as more and more smaller scales were
resolved in the simulation. Recently, simulation results of \cite{HRY16}
were combined with some more simulations along the lines of the dynamo
model of \cite{FF14} to calculate the values of $\alpha_\mathrm{K}$
and $\alpha_\mathrm{M}$ and interpret the behavior of the magnetic
energy as being due to small-scale magnetic helicity driving (Mei Zhang,
private communication).

Obviously, our approach does no longer derive from first principles,
unlike standard mean-field dynamo theory which is a rigorous theory within
the regime of applicability (e.g., for small magnetic Reynolds numbers).
Nevertheless, we feel that in practice, when the turbulent eddies are
no longer small compared with the scale over which the mean field varies,
our approach is more realistic than
the conventional one where nonlocality is neglected altogether.
It is not obvious whether nonlocality is able to produce decisively
different solutions compared with the usual dynamo solutions.
It would be important to explore a wider range of models, aiming for
realistic setups applied not only to the Sun, but also to galactic
and accretions disc dynamos for which a number of three-dimensional
simulations are already available.

One of the surprising results of global convectively-driven dynamo
simulations is the occurrence of equatorward migration in the nonlinearly
saturated regime in direct numerical simulations that has not yet been
explained (K\"apyl\"a et al.\ 2012; Augustson et al.\ 2015;
Strugarek et al.\ 2017).
A possible explanation for this behavior might be the occurrence of
a secondary dynamo wave driven by current helicity in the nonlinear regime.
To model this phenomenon correctly, nonlocality in space in necessary
to prevent the formation of artificial small-scale structures.
However, subsequent work of Warnecke et al.\ (2014) demonstrated that
the cause of the equatorward migration is actually a peculiar feature
in the differential rotation profile at intermediate latitudes, which
they referred to as ``tongues'' and that the migration direction follows
simply from the Parker--Yoshimura rule (Parker 1955, Yoshimura 1975).
On the other hand, Augustson et al.\ (2015) proposed that the equatorward
migration is not related to peculiar features in the differential
rotation, but that it is actually a consequence of nonlinearity.

It would be useful to allow for more realistic flows that include
meridional circulations and differential rotation generated in a more
self-consistent manner using the $\Lambda$ effect (R\"udiger 1980, 1989),
i.e., a mean-field parameterization of the Reynolds stress in the averaged
momentum equation; see Brandenburg et al.\ (1992) and Rempel (2005)
for models in that direction.
Again, nonlocality should not be neglected here either.
Indeed, it appears now natural to apply a similar procedure and replace
the usual parameterization of the Reynolds stress by an evolution
equation with a diffusion term, similar to \Eq{NonlocalEvol}.
The same applies also to stellar mixing length theory where such
a treatment of nonlocality would seem to be able to deal in a natural
way with convective overshoot, for example.

\acknowledgements
This research was supported in part by the NSF Astronomy and Astrophysics
Grants Program (grant 1615100), and the University of Colorado through
its support of the George Ellery Hale visiting faculty appointment.
We acknowledge the allocation of computing resources provided by the
Swedish National Allocations Committee at the Center for Parallel
Computers at the Royal Institute of Technology in Stockholm.
This work utilized the Janus supercomputer, which is supported by the
National Science Foundation (award number CNS-0821794), the University
of Colorado Boulder, the University of Colorado Denver, and the National
Center for Atmospheric Research. The Janus supercomputer is operated by
the University of Colorado Boulder.
The input files as well as some of the output files of the
simulation are available under
\url{http://www.nordita.org/~brandenb/projects/spherical-geom}.



\begin{thebibliography}{}

\bibitem[Augustson et al.(2015)]{ABMT15}
Augustson, K., Brun, A. S., Miesch, M., \& Toomre, J.\yapj{2015}{809}{149}

\bibitem[Blackman \& Field(2002)]{BF02}
Blackman, E. G., \& Field, G. B.\yprl{2002}{89}{265007}

\bibitem[Blackman \& Field(2003)]{BF03}
Blackman, E. G., \& Field, G. B.\ypf{2003}{15}{L73}

\bibitem[Brandenburg(2016)]{Bra16}
Brandenburg A.\yapj{2016}{832}{6}

\bibitem[Brandenburg et al.(2004)]{BKM04}
Brandenburg, A., K\"apyl\"a, P., \& Mohammed, A.\ypf{2004}{16}{1020}

\bibitem[Brandenburg et al.(1992)]{BMT92}
Brandenburg, A., Moss, D., \& Tuominen, I.\yana{1992}{265}{328}

\bibitem[Brandenburg et al.(2008a)]{BRS08}
Brandenburg, A., R\"adler, K.-H., \& Schrinner, M.\yana{2008a}{482}{739}

\bibitem[Brandenburg \& Subramanian(2005a)]{BS05}
Brandenburg, A., \& Subramanian, K.\yjour{2005a}{Phys.\ Rep.}{417}{1}

\bibitem[Brandenburg \& Subramanian(2005b)]{BS05c}
Brandenburg, A., \& Subramanian, K.\yan{2005b}{326}{400}

\bibitem[Brandenburg et al.(2009)]{BSV09}
Brandenburg, A., Svedin, A., \& Vasil, G. M.\ymn{2009}{395}{1599}

\bibitem[Brandenburg \& Tuominen(1988)]{BT88}
Brandenburg, A., \& Tuominen, I.\yjour{1988}{Adv. Space Sci.}{8}{185}
{189}{Variation of magnetic fields and flows during the solar cycle}

\bibitem[Canuto \& Mazzitelli(1991)]{CM91}
Canuto, V. M., \& Mazzitelli, I.\yapj{1991}{370}{295}

\bibitem[Chatterjee et al.(2011)]{CGB11}
Chatterjee, P., Guerrero, G., \& Brandenburg, A. \yana{2011}{525}{A5
}

\bibitem[Chatterjee et al.(2004)]{CNC04}
Chatterjee, P., Nandy, D., \& Choudhuri, A. R.\yana{2004}{427}{1019}

\bibitem[Choudhuri et al.(1995)]{CSD95}
Choudhuri, A. R., Sch\"ussler, M., \& Dikpati, M.\yana{1995}{303}{L29}

\bibitem[Del Sordo et al.(2013)]{DSGB13}
Del Sordo, F., Guerrero, G., \& Brandenburg, A.\ymn{2013}{429}{1686}

\bibitem[Dikpati \& Charbonneau(1999)]{DC99}
Dikpati, M., \& Charbonneau, P.\yapj{1999}{518}{508}

\bibitem[Fan \& Fang (2014)]{FF14}
Fan, Y. \& Fang, F. \yapj{2014}{789}{35}

\bibitem[Hotta et al.\ (2016)]{HRY16}
Hotta, H., Rempel, M., \& Yokoyama, T.\ysci{2016}{351}{1427}

\bibitem[Hubbard \& Brandenburg(2009)]{HB09}
Hubbard, A., \& Brandenburg, A.\yapj{2009}{706}{712}

\bibitem[Hubbard \& Brandenburg(2011)]{HB11}
Hubbard, A., \& Brandenburg, A.\yapj{2011}{727}{11}

\bibitem[Hubbard \& Brandenburg(2012)]{HB12}
Hubbard, A., \& Brandenburg, A.\yapj{2012}{748}{51}

\bibitem[K\"apyl\"a et al.(2012)]{KMB12}
K\"apyl\"a, P. J., Mantere, M. J., \& Brandenburg, A.\yapjl{2012}{755}{L22}

\bibitem[Krause \& R\"adler(1980)]{KR80}
Krause, F., \& R\"adler, K.-H.\ybook{1980}
{Mean-field Magneto\-hydro\-dy\-na\-mics and Dynamo Theory}{Oxford: Pergamon Press}

\bibitem[Krivodubskii(1984)]{Kri84}
Krivodubskii, V. N.\ysov{1984}{28}{205}

\bibitem[Madarassy \& Brandenburg(2010)]{MB10}
Madarassy, E. J. M., \& Brandenburg, A.\ypre{2010}{82}{016304}

\bibitem[Mitra et al.(2009)]{Mitra09}
Mitra, D., K\"apyl\"a, P. J., Tavakol, R., \& Brandenburg, A.\yana{2009}{495}{1}

\bibitem[Parker(1955)]{Par55}
Parker, E. N.\yapj{1955}{122}{293}

\bibitem[Parker(1979)]{Par79}
Parker, E. N.\ybook{1979}{Cosmical magnetic fields}{Clarendon Press, Oxford}

\bibitem[Pipin(2008)]{Pip08}
Pipin, V. V.\ygafd{2008}{102}{21}

\bibitem[Pipin(2017)]{Pip17}
Pipin, V. V.\ymn{2017}{466}{3007}

\bibitem[Pipin \& Kosovichev(2011)]{Pipin}
Pipin, V. V., \& Kosovichev, A. G.\yapj{2011}{727}{L45}

\bibitem[R\"adler(1976)]{Rae76}
R\"adler, K.-H.\yproc{1976}{323}
{Basic Mechanisms of Solar Activity, Proceedings from IAU Symposium
No.\ {\bf 71} held in Prague, Czechoslovakia}
{V.\ Bumba and J.\ Kleczek}{D.\ Reidel Publishing Company Dordrecht}

\bibitem[R\"adler(1986)]{Rae86}
R\"adler, K.-H.\yan{1986}{307}{89}

\bibitem[Rempel(2005)]{Rem05}
Rempel, M.\yapj{2005}{622}{1320}

\bibitem[Rheinhardt \& Brandenburg(2012)]{RB12}
Rheinhardt, M., \& Brandenburg, A.\yan{2012}{333}{71}

\bibitem[R\"udiger(1980)]{Rue80}
R\"udiger, G.\ygafd{1980}{16}{239}

\bibitem[R\"udiger(1989)]{Rue89}
R\"udiger, G.\ybook{1989}{Differential rotation and stellar convection:
Sun and solar-type stars}{Gordon \& Breach, New York}

\bibitem[R\"udiger \& Hollerbach(2004)]{RH04}
R\"udiger, G., \& Hollerbach, R.\ybook{2004}{The magnetic universe}{New York: Wiley-VCH, Weinheim}

\bibitem[Stix(2002)]{Sti02}
Stix, M.\ybook{2002}{The Sun: An introduction}
{Springer-Verlag, Berlin}

\bibitem[Strugarek et al.(2017)]{SBCBdN17}
Strugarek, A., Beaudoin, P., Charbonneau, P., Brun, A. S., \& do Nascimento, J.-D.\ysci{2017}{357}{185}

\bibitem[Subramanian \& Brandenburg(2006)]{SB06}
Subramanian, K., \& Brandenburg, A.\yapj{2006}{648}{L71}

\bibitem[Sur et al.(2008)]{SBS08}
Sur, S., Brandenburg, A., \& Subramanian, K.\ymn{2008}{385}{L15}

\bibitem[Vishniac \& Cho(2001)]{VC01}
Vishniac, E. T., \& Cho, J.\yapj{2001}{550}{752}

\bibitem[Vitense(1953)]{Vit53}
Vitense, E.\yjour{1953}{Z.\ Astrophys.}{32}{135}

\bibitem[Warnecke et al.(2014)]{WKKB14}
Warnecke, J., K\"apyl\"a, P. J., K\"apyl\"a, M. J., \& Brandenburg, A.\yapjl{2014}{796}{L12}

\bibitem[Yoshimura(1975)]{Yos75}
Yoshimura, H.\yapjs{1975}{29}{467}

\end{thebibliography}
\end{document}